\documentclass[referee]{aa} % for a referee version
\usepackage{graphicx}
\usepackage{txfonts}
\usepackage{natbib}
\bibpunct{(}{)}{;}{a}{}{,} % to follow the A&A style
\begin{document}
   \title{The end of super AGB and massive AGB stars}

   \subtitle{I. The instabilities that determine the final mass of AGB stars}

   \author{H.H.B. Lau
          \inst{1,3},
          P. Gil--Pons
	  \inst{2,3},
          C.Doherty,
          \inst{3}
          \and
          J.Lattanzio\inst{3}
%\fnmsep\thanks{Just to show the usage
%          of the elements in the author field}
          }

   \institute{ Argelander Institute for Astronomy, University of Bonn, Auf dem Huegel 71, D-53121 Bonn, Germany,  
              \email{h.lau@monash.edu}
         \and
             Department of Applied Physics, Polytechnical University of Catalonia, 08860 Barcelona, Spain
         \and
             Monash Centre for Astrophysics, School of Mathematical Sciences, Monash University, Victoria 3800, Australia
             }
   \date{Received; accepted}

% \abstract{}{}{}{}{} 
% 5 {} token are mandatory
 
  \abstract
   {The literature is rich in analysis and results related to thermally pulsing-asymptotic giant branch (TP-AGB) stars, but the problem of the instabilities that arise and cause the divergence of models during the late stages of their evolution is rarely addressed.
   }
   {We investigate the physical conditions, causes and consequences of the interruption in the calculations of massive AGB stars in the late thermally-pulsing AGB phase.
   }
   {We have thoroughly analysed the physical structure of a solar metallicity $8.5 M_{\odot}$  star and described the physical conditions at the base of the convective envelope (BCE) just prior to divergence.
   }
   {We find that the local opacity maximum caused by M-shell electrons of Fe-group 
elements lead to the accumulation of an energy excess, to the departure of 
thermal equilibrium conditions at the base of the convective envelope and,
eventually, to the divergence of the computed models. For the $8.5 M_{\odot}$ 
case we present in this work the divergence occurs when the envelope mass is 
about $2 M_{\odot}$. The remaining envelope masses range
 between  somewhat less than 1 and more than $2 M_{\odot}$ for stars with initial masses between 
7 and $10 M_{\odot}$ and, therefore, our results are relevant for the evolution
and yields of super-AGB stars. 
If the envelope is ejected as a consequence of the instability we are considering, the occurrence of electron-capture supernovae would be avoided at solar metallicity.
   }
%The envelope is likely to be ejected as a consequence of the instability, and hence electron-capture supernovae are very unlikely to occur at solar metallicity
   {}

\authorrunning{H.H.B.Lau, P.Gil--Pons, C.Doherty, J.Lattanzio}
\titlerunning{The end of super and massive AGB stars}
\maketitle
%
%________________________________________________________________

\section{Introduction}

One of the most complicated phases of the evolution of low- and intermediate-
 mass stars is the thermally pulsing-asymptotic giant branch (TP-AGB) stage. The physical conditions
that are sampled by models of this phase are quite extreme and the recurring thermal
instability of the helium-burning shell requires codes to be robust on
short timescales during the thermal pulse cycle. It is common for convergence problems
to arise during the study of such stars, although rarely are these problems
discussed in the recent literature.
The failure to converge usually occurs during the later
thermal pulses, when the envelope mass is relatively low due to mass loss through the AGB phase. We are currently using the Monash stellar
evolution code \textsc{MONSTAR} \citep{Cam08} to study the evolution of
super asymptotic giant branch (super-AGB) stars and have
encountered such a problem.

When the envelope mass is below $\approx$ $2~M_\odot$, the code's solution typically tends to a configuration with a negative gas pressure
at the bottom of the convective envelope. We find that this
convergence problem can be delayed by slowly increasing $\alpha$,
the ratio of the convective mixing length to the pressure scale height. Our models are originally evolved with $\alpha =1.75$.
Typically, with a slightly higher $\alpha$, a few more thermal pulses could be modelled until the same convergence problem occurs again.
This works until the envelope mass is in the region of  $0.5-1 M_\odot$, when increasing $\alpha$ will no longer avoid the problem.
A typical value of $\alpha\sim 4$ is needed to evolve the model up to this
stage. Previously, \cite{Herwig2001} has also suggested that a
  higher $\alpha$ value (of 3) helps to ease post-AGB numerical
  simulations. Recipes of adopting higher values of $\alpha$ during the
  later 
stages of the AGB evolution are also mentioned in \cite{MillerBertolami2006}, \cite{Kitsikis2008} and \cite{Weiss2009}.

In this paper we seek a physical explanation for
such behaviour and find that the radiation pressure is so
large that it supplies all the pressure required for
hydrostatic equilibrium. In other words,  the local luminosity
exceeds the Eddington limit.  This type of instability has
previously been discussed by \cite{woo86} who suggested that
the envelope could be ejected and the star would then evolve
to the planetary nebula. Eddington limit luminosities as the cause
  of the numerical problems during low-mass envelope thermal pulses has
  already been identified by \cite{Lawlor2003}\footnote{Note that Table 4
    in this paper is in error, and that all entries for $M\ge 4 M_\odot$
    should be in italics  (\citealt{Lawlor2006}; MacDonald, private communication)} and \cite{MillerBertolami2006a}.

The stability of stellar models during the late thermally pulsing AGB phase has also been 
considered by \cite{han94}, \cite{wag94}, \cite{woo93}, \cite{ren92}, \cite{tuc84}, \cite{bar80}, \cite{tuc78}, \cite{woo73}, and \cite{pac68}.
The transition from AGB to post-AGB evolution has been considered,
from the observational point of view, by \cite{Eng09}.

In this paper, we discuss in detail the cause of the
instability in the models and whether real stars do encounter
such an instability in nature. We also discuss possible
scenarios for when this instability may occur and try to
determine whether the envelope would be ejected based on
the results of our models.
The next section of the present work gives a physical description of
the instability, considering the cases when energy transport is either 
dominated by radiation or cases when convection is efficient.
In the third section a hypothesis for the cause of 
the instability based on the existence of an iron element opacity 
peak is proposed and tested. In the fourth section we consider 
the possible fate of the star after the instability. Finally we present the main conclusions and 
discuss our work.

\section{Physical description of the instability}
\begin{figure}
%   \vspace{2.0cm}
   \includegraphics{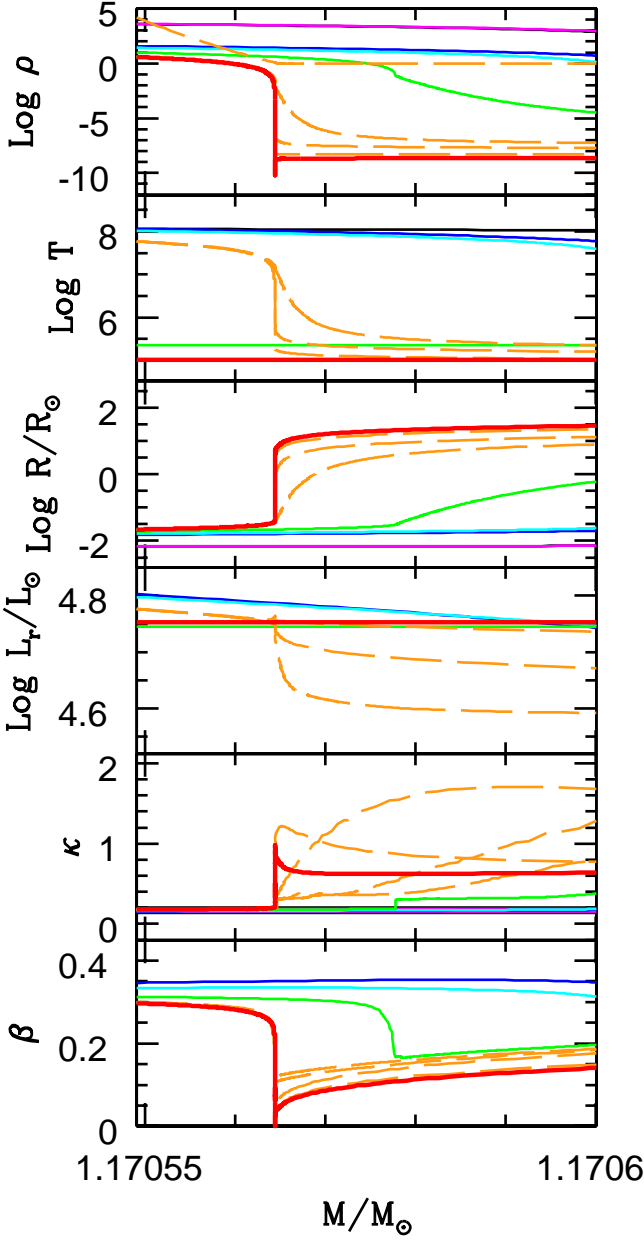}
\caption{Density, temperature, radius, luminosity, $\kappa$ and $\beta$, the ratio of gas pressure to total pressure in the instability region. The model profiles represented in solid lines correspond to the times labelled in Fig~\ref{fig:lastlumconv} and in particular thick lines represent the last converged models. Dashed lines represent the final progress of the instability.}
\label{fig:rhotkappa}
  \end{figure}

The computation of the late stages of massive TP-(super)AGBs is hampered by the occurrence of a type of instability that causes evolutionary models to fail to converge and can affect our understanding of the most advanced phases of the evolution and eventually the resulting white dwarf remnants. 
This instability is characterised by the development of a zone at the base of
the convective envelope in which the density and the gas pressure, $P_{\mathrm{gas}}$
 tend to zero
and, therefore, only radiation contributes to the total pressure in the
equation of state. Eventually the models converge to solutions with local
$P_{\mathrm{gas}}$ below zero near the base of the convective envelope (BCE) and the 
codes finally crash. As we will show in the next subsection, this is equivalent
 to having local luminosity values higher than the Eddington luminosity.
The temperature in the stellar zone in which the instability develops
{\bf is} typically $1-3\times 10^5\,$K and thus much higher than the values found in the region where hydrogen and helium are
partially ionized. Hence the instability we are concerned with is not
directly related to the instability described by \cite{wag94}. These authors identified
the local decrease in efficiency of the energy transport due to the
recombination of hydrogen as the main cause of the instability that quenched
the computation of AGB evolution.

We have noticed the presence of the instability described above
 when using the Mount Stromlo and
Monash Stellar Evolution code (MONSTAR), but from Siess (private communication) we know that the stellar evolution code STAREVOL \citep{Sie10}, presents 
the same convergence problems during late massive AGB and super-AGB evolution. Both codes and their respective super-AGB models have been described by \cite{Doh10}.
It is worth noting that such convergence problems are common. Many evolution codes experience numerical problems that could be related to this instability. The MONSTAR code \citep{Cam08} and the STAREVOL code \citep{Sie07,Sie10} and the STARS code \citep{Sta07,lau09}
have convergence problems during the late massive AGB or super-AGB evolution.

We have also checked that whether this instability occurs is insensitive to most input physics of the code. For example, we have used different implementations of the 
OPAL opacities, of the treatment of mixing (with or without time independent mixing), or of the mass-loss prescription: \cite{Rei78}, \cite{Vas93},
or \cite{Blo95}. The models presented here are modelled with the \cite{Vas93} mass-loss prescription.
These make only minor differences to the occurrence of the
instability. Unless one
employs a very high mass-loss rate, such as the one used by \cite{Kovetz2009}, the instability cannot be avoided.\footnote{For their massive stars, \cite{Kovetz2009} adopted \cite{Blo95} mass loss presciption with $\eta=3$.}
In fact different input physics does change the envelope mass when the instability begins, but they can not prevent it from occuring.
The opacity calculation could on the other hand, have a huge effect on whether this instability occurs, as shown in a later section. Byremoving the Fe opacity peak, it is possible to avoid the instability. Computations with old Los Alamos(LAOL) opacities may not suffer this instability e.g. \cite{Bloecker1995}. In fact, the main improvement of the new opacities (OPAL/OP) tables is the treatment of Fe and iron-group opacities in the range of temperatures $10^6-10^5$ K \citep{Iglesias1996}. This could explain why instabilities are more common with the adoption of new opacities. This is consistent with our identification of the iron opacity peak as the cause of the instability in the models discussed in this paper.

In terms of numerical details, we have tried varying the obvious things, such as
increasing the temporal or spatial resolution, but these make little difference. We may delay the instability for a few pulses but not eliminate it.
However, we discovered that increasing the mixing length parameter $\alpha$
enables the code to continue to converge for a little longer. This was also discovered by \cite{Herwig2001}, \cite{MillerBertolami2006}, \cite{Kitsikis2008} and \cite{Weiss2009}. In this way we can push the
evolution for a few more thermal pulses, although inevitably the
convergence problem re-appears. A further increase in $\alpha$
again delays the problem. This is a hint to the underlying physics but
we are not free to increase $\alpha$ arbitrarily and without limit!

\begin{figure}[t]
   \centering
%   \vspace{2.0cm}
   \includegraphics[scale=0.55,angle=0]{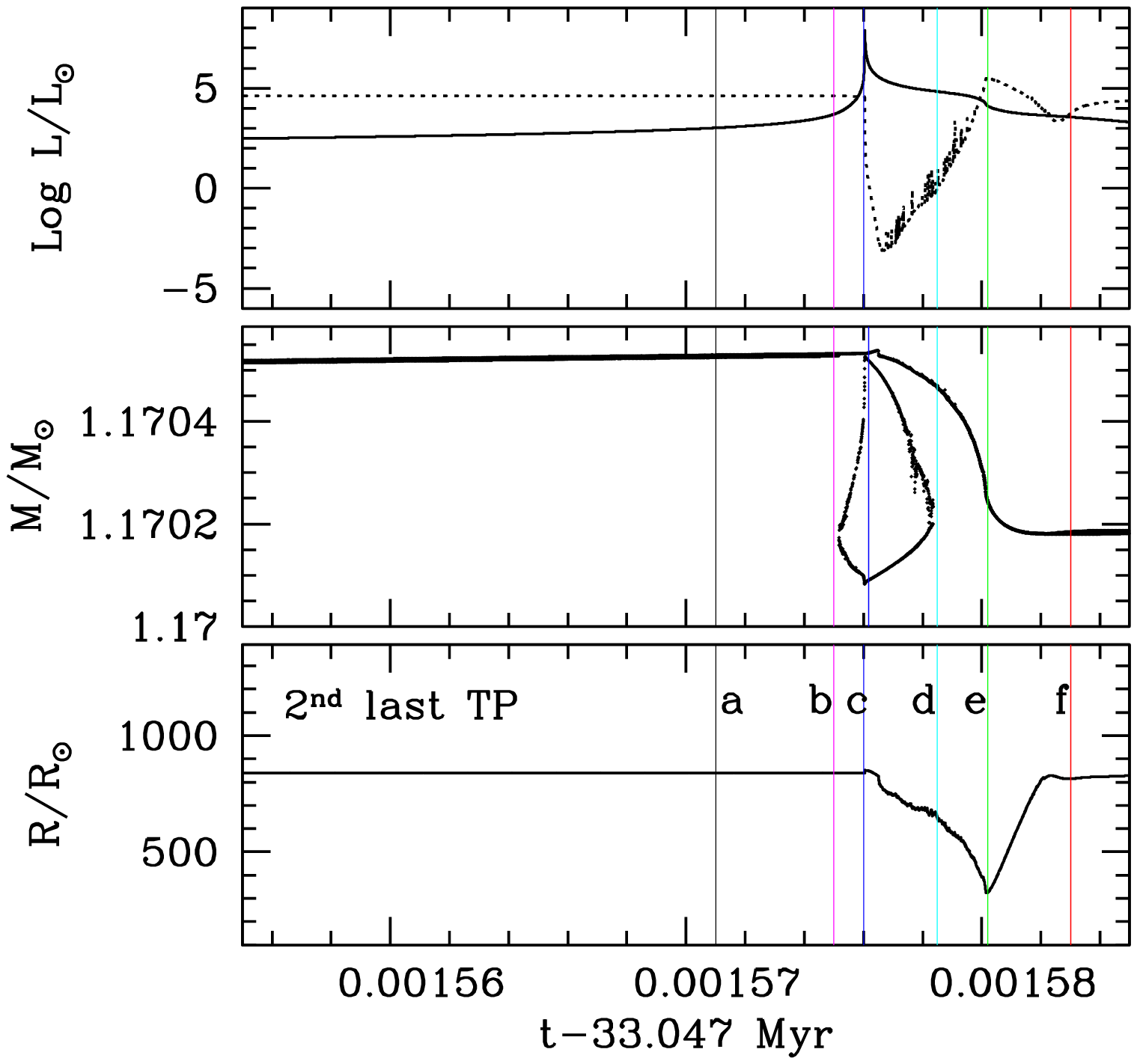}
   \caption{Upper panel: evolution of luminosity associated to H-burning
 (dotted line) and He-burning (solid line); middle panel: evolution of convection; lower panel: evolution of the surface radius, all during the second last thermal pulse. Relevant structure profiles of models labelled a to f appear in Fig.~\ref{fig:binden2}.}
\label{fig:2lastlumconv}
   \end{figure}

\begin{figure}[t]
   \centering
%   \vspace{2.0cm}
   \includegraphics[scale=0.55,angle=0]{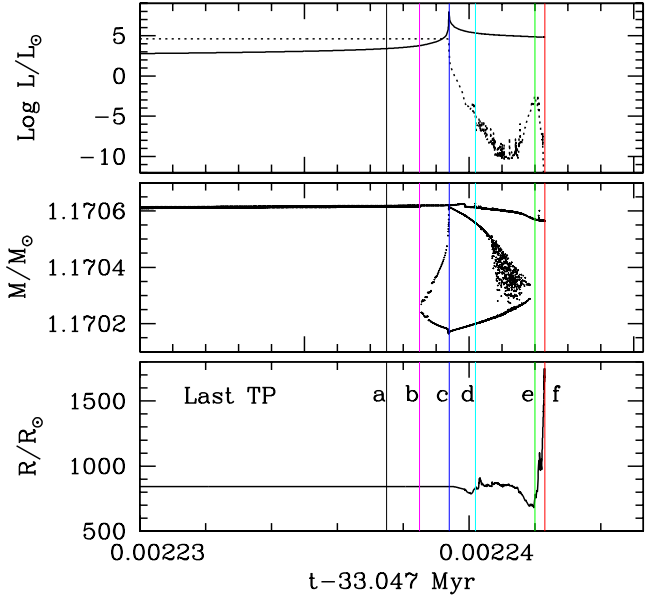}
   \caption{Upper panel: evolution of luminosity associated to H-burning
 (dotted line) and He-burning (solid line); middle panel: evolution of convection; lower panel: evolution of the surface radius, all during the last thermal pulse. Relevant structure profiles of models labelled a to f appear in Fig.~\ref{fig:binden}.}
\label{fig:lastlumconv}
   \end{figure}

The instability is manifested in the mesh points at the base of the convective envelope. Physical
conditions at this time are shown in Fig.~\ref{fig:rhotkappa} to \ref{fig:leddiv} for a $8.5 M_{\odot}$ star of solar metallicity. The MONSTAR code uses temperature and pressure as dependent
variables and the iterations soon produce a T and P corresponding to such a high
radiation pressure, $P_{\mathrm{rad}}$ that the implied gas pressure $P_{\mathrm{gas}}=P-P_{\mathrm{r}}$is negative, as in indicated in Fig.~\ref{fig:rhotkappa}.
From Fig.~\ref{fig:lednodiv} we can see that the instability starts at the boundary between radiative and convective zones. As the instability develops, the density at the bottom of the convective envelope decreases (Fig.~\ref{fig:rhotkappa}) because the model star is expanding,
as evidenced by the rapid increase in radius around that region (figure~\ref{fig:rhotkappa}).

% Also,  the instability occurs at the bottom of the envelope, with much higher temperature than the region where He and H get ionized. The instability is not related to the instability described by \cite{Wagenhuber94}  where they attribute these problems to H recombination. Also,it is worth noting that neither switching off mass loss nor increasing mass loss rate do not help the models to converge.

%Both Monash code and Siess's code have convergence problem in massive AGB stars or SAGB stars for the meshpoints at the base of the convective envelope. Typically, the codes converges into a models with gas pressure below zero.

%Usually, by increasing the mixing length $\alpha$, the code will be able to converge for several thermal pulses. Also,  the instability occurs at the bottom of the envelope, with much higher temperature than the region where He and H get ionized. The instability is not related to the instability described by \cite{Wagenhuber94}  where they attribute these problems to H recombination. Also,it is worth noting that neither switching off mass loss nor increasing mass loss rate do not help the models to converge.

%%%%%%%%%%%%%%%%%%%%%%%%%%%%%%%%%%%%%%%%%%%%%%%%%%%%%%%%%%%%%%%%%%%%%%%%%%%%

%%%%%%%%%%%%%%%%%%%%%%%%%%%%%%%%%%%%%%%%%%%%%%%%%%%%%%%%%%%%%%%%%%%%%%%%%%%%

\cite{woo86}, section IIIb, while developing an analysis of hydrostatic sequences
of planetary nebulae, justified the existence of the TP-AGB instability.
They computed models that would produce the nuclei of the PNe between 0.60 and $0.89 M_{\odot}$, and realised that, at some point in the late evolution of TP-AGB stars,
the radiation pressure, $P_{\rm rad}$, becomes so large that its contribution accounts
practically for the total pressure at the base of the convective envelope. So
$P_{\rm gas}$ is forced to be zero. These authors suggest that, when this phenomenon
takes place in a real star, a hydrodynamic expansion of the stellar envelope occurs and that convergence problems develop in the models as a consequence of the lack of existence of a hydrostatic solution.

%According to \cite{woo86} section III b, the radiation pressure in some models becomes so large in the envelope, that it
%supplies all the pressure support required by the model, and thus the gas
%pressure is forced to zero. They suggested in a real star, when it
%happens, a hydrodynamic expansion of the envelope of the star occurs in a
%real star, and the convergence problem results from the lack of existence
%of hydrostatic solution.

This is consistent with the instability we are studying in this work. In particular, $\beta$, the ratio of gas pressure to total pressure at the base of envelope, drops as the stars evolve along the last thermal pulses, with our last successfully converged models yielding values of $\beta$ below 0.001.
%when the model fails to converge and beta
%could be below 0.001 in previous successfully converged models.
The behaviour of $P_{\rm gas}$ tending to (or falling below) zero
is accompanied and, in fact, is equivalent to local luminosity
values reaching (or exceeding) the Eddington luminosity.
We will develop this further in the next subsections.

%Also, the local luminosity at the bottom of the convective envelope keeps increaseing as the star evolves, and eventually reaches
%99.5\% of the local Eddington luminosity before the code crashes.

%In fact, exceeding the local Eddington luminosity is equivalent to $\beta$ falls below zero, as shown in the next subsection.
%%%%%%%%%%%%%%%%%%%%%%%%%%%%%%%%%%
%%%%%%%%%%%%%%%%%%%%%%%%%%%%%%%%%%%%%%%%%%%%%%%%%%%%%%%%%%%%%%%%%%%%%%%%%%%%%%%%%%%

\subsection{The importance of the convective efficiency}
The case where the energy is carried by radiation alone has already been presented in \cite{woo86}
The situation is more complicated when convection is
present because it adds a second energy transport mechanism. The case
of vanishingly low convective efficiency corresponds to the radiative
case. We are interested in how convective
efficiency modifies the behaviour of the star.

In a region where convection is efficient, the local luminosity
can exceed the Eddington limit without encountering the above instability.
 This occurs frequently during the lives of stars. The core helium flash is
one example, where the luminosity can be $10^8 L_\odot$ or more, greatly
exceeding the local Eddington value. Instead convection sets in when the local luminosity exceeds the Eddington limit and the highly efficient convection
carries most of the energy and the instability does not develop. Similarly,
during AGB thermal pulses themselves, in the helium burning region we have
similarly high values and again the efficient convection helps to keep $\beta$ positive even though the local luminosity exceeds the Eddington value.

In a region where convection is efficient, most of the energy is transported
by convection instead of radiation. We derive below a corresponding
Eddington Limit for a convective region and show that the value
depends on the efficiency of convection. In regions where convection is
inefficient the local Eddington limit still applies because most
energy is transported by radiation.

We begin with the equation of pressure equilibrium.
\begin{equation}
\frac{d P}{d r} = -{{\rho G M_{\mathrm {r}}}\over{r^2}}
\end{equation}

where, $P$ is the pressure, $r$ is the radius, G is the gravitational constant and $M_{\mathrm {r}}$ is the mass within the radius,
and from energy transport by radiation.

The energy transfer equation is given by
\\
\noindent
\begin{equation}
\frac {d T}{d r}= \frac {G M_{r} T}{4\pi r^4P}\nabla
\end{equation}
where $T$ is the temperature, $\kappa$ is the opacity, $\rho$ is the density and $L_{\mathrm {r}}$ is the local luminosity. In the radiative case, $\nabla=\nabla_{\mathrm {r}}$ and $\nabla_{\mathrm {r}}$ is defined as $\frac {3}{16 \pi a c G} \frac {\kappa L_{\mathrm {r}} P} {m T^4}$.

The above equation can be rewritten in the form
\noindent
\begin{equation}
\frac {d T}{d r}=\frac {-3}{16 \pi a c} \frac {\kappa \rho L_{\mathrm {r}}} {r^2 T^3} \left(\frac {\nabla} {\nabla_{\mathrm {r}}}\right)
\end{equation}
where
\noindent
\begin{equation}
\nabla_{\mathrm {r}}= \frac {3}{16 \pi a c G} \frac {\kappa L_{\mathrm {r}} P} {m T^4}
\end{equation}
is the gradient if the star is radiative and $\nabla$ is the actual gradient of the star.
By definition the radiative pressure $P_{\rm rad}= \frac{1}{3} a T^4$, so we can obtain
\noindent
\begin{equation}
\frac{dP_{\rm rad}}{dr}= \frac {\kappa\rho L_{\mathrm {r}}}{4\pi c r^2} \left(\frac {\nabla} {\nabla_{\mathrm {r}}}\right)
\end{equation}

Combining with $\frac{d P}{d r} = -{{\rho G M_{\mathrm {r}}}\over{r^2}}$ we obtain the differential equation
\noindent
\begin{equation}
\frac{dP_{\rm rad}}{d P} = \frac {\kappa L_{\mathrm {r}}}{4\pi c G M_{\mathrm {r}}} \left(\frac {\nabla} {\nabla_{\mathrm {r}}}\right)
\end{equation}

As we can see from the above equation, a jump in local luminosity or opacity will cause the local radiation pressure to increase much more rapidly than the local pressure. It is therefore possible that the radiation pressure completely dominates the total pressure.

In particular, for the base of the convective envelope (BCE), the stellar region in which we are interested, $L_{\mathrm {r}}$ and $M_{\mathrm {r}}$ can be
well approximated by the total luminosity of the star, $L$, and its core mass $M_c$, respectively. If we assume the envelope mass is small and take $M_{\mathrm {r}}$ as the core mass $M$, then the luminosity $L$ is a constant and equals to the value at the base of the convective envelope. By integrating over a narrow region around the BCE and using a local average value for
the opacity $<\kappa>$, we can express the ratio of radiation pressure to total pressure at the bottom of the envelope
as follows:
\noindent
\begin{equation}
\frac{P_{\rm rad}}{P}=1-\beta \approx \frac {<\kappa> L }{4\pi c G M_c} \left(\frac {\nabla} {\nabla_{\mathrm {r}}}\right) = \frac {L} {L_{\rm Edd}} \frac{\nabla}{\nabla_{\mathrm {r}}}= \frac {L} {L'_{\rm Edd}}
\end{equation}
where ${<\kappa>} = P^{-1}\int_0^P \kappa dP$, which is the average of the opacity over the region.

From the above equation, we note that $\beta <0$ is equivalent to $L > L'_{\rm Edd}$, where $L'_{\rm Edd}= {L_{\rm Edd}} \frac{\nabla_{\mathrm {r}}}{\nabla}$. In convective region where convection is efficient $\nabla$ is very nearly equal to the adiabatic gradient ($\nabla_a$) and is much smaller than $\nabla_{\mathrm {r}}$, and hence the equivalent Eddington limit is much larger and the radiation pressure does not dominate the total pressure. However, in regions where convection is inefficient, the value of $\nabla$ approaches the value of $\nabla_{\mathrm {r}}$, and the above equation approaches the equation $\frac{P_{\rm rad}}{P}=1-\beta \approx \frac {<\kappa> L }{4\pi c G M_{\rm c}}= \frac {L} {L_{\mathrm{Edd}}}$, that is, the Eddington limit converges back to the classical value in a radiative region.

The actual value $\nabla$ depends on the efficiency of the convection and must satisfy the relation $\nabla_{\mathrm {r}}>\nabla>\nabla_{\rm a}$, where $\nabla_{\rm a}$ is the adiabatic gradient.
But it is important to keep in mind that radiation and convection can coexist in the same region of a star and both can be relatively
efficient. It is only in some particular regions, such as the stellar centre, in which convection fully dominates and
radiation can be ignored.

%During the motion for the elements during convection, it is still transporting energy by radiation. For example, in the center of the star, the energy transported by radiation can be ignored because convection is very efficient, but this is not the case everywhere in the star.

\cite{Kippenhahn} (page 51) derive the value of $\nabla$ based on mixing length theory. The most important quantity is
\noindent
\begin{equation}
U= \frac {3acT^3}{C_{\rm P} \rho^2 \kappa l_{\rm m}^2} \sqrt{\frac{8H_{\rm P}}{g\delta}}
\end{equation}
where $l_{\rm m}$ is the mixing length and $H_P$ is the pressure scale-height. The quantity U can be taken as the ratio of the radiative 'conductivity' to the convective 'conductivity'. If U is large, convection is ineffective and cannot transport a substantial fraction of the luminosity. From the equation, also in \cite{Kippenhahn}
\begin{equation}
(\nabla-\nabla_{\rm e})^{3/2}= \frac {8}{9} U (\nabla_{\mathrm {r}}-\nabla)
\end{equation}
where $\nabla_{\rm e}$ is related to the temperature changes of the moving elements. We see that, in order for the equation to have solution when $U$ is larger, $\nabla \approx \nabla_{\mathrm {r}}$ and hence most of the energy is transported by radiation.

Because U is inversely proportional to $\rho^2$, convection is very
efficient in the dense central part of the star but not necessarily
in the envelope where the density is low. In our model, U is of the order of 100 at the bottom of
the convective envelope, and according to \cite{Kippenhahn} this value
of U implies radiation dominates over convection for the energy transport.
Therefore, $\nabla \sim \nabla_{\rm r}$ and the actual Eddington limit is close
to the radiative value. When the local luminosity exceeds
this the instability occurs.

Notice that this naturally gives us the explanation for
why increasing the mixing-length can delay the onset of
the convergence problems. We see that U is inversely proportional
to $l_{\rm m}^2$. So by increasing $\alpha$ we increase the value of $l_{\rm m}$,
decreasing the
value of U, and forcing the convection to be more efficient.
A smaller U implies more energy is transported by convection.
Hence $\nabla$ decreases and the effective Eddington luminosity increases
beyond the radiative value. Thus by increasing $\alpha$ we enable more
efficient convection to carry the energy away and the instability is avoided.

It is interesting to see how the above argument is reflected on an actual computation of a
full evolutionary sequence. The next few paragraphs compare our previous analysis with the results of 
an actual computation of the final stages of a TP-(super)AGB star. We have considered a
$8.5 M_{\odot}$ star of solar metallicity, from its main sequence, until the instability is reached 
after 152 thermal pulses. 
At this time our model star has $2.71 M_{\odot}$, of which $1.17 M_{\odot}$ correspond to the core 
and the remaining $1.54 M_{\odot}$ constitute the envelope.

Fig.~\ref{fig:2lastlumconv} shows the temporal evolution of luminosity, convection and radius
during the second last thermal pulse. The
upper panel shows the temporal evolution of the luminosity associated to H-burning, $L_{\rm H}$, and He-burning, $L_{\rm He}$,
the middle panel shows the evolution of the BCE and the convective shell associated with the He-flash
and the lower panel shows the variation of the surface radius with time. The labelled vertical lines correspond
 to selected models whose structure profiles are shown in fig~\ref{fig:binden2} and ~\ref{fig:binden}.

Fig.~\ref{fig:lastlumconv} shows the same as the former figure, for the case of the last thermal
pulse. As we can see, when $L_{\rm He}$ decays after the flash $L_{\rm H}$ increases and tends to recover
its interpulse value. But, opposite to what happens in former thermal-pulses, $L_{\rm H}$ reaches
a maximum of about $10^{-3} L_{\odot}$ and then decreases again. This is because the stars tries to expand around the region where the gas pressure 
{\bf is} approaching zero.
As a result, the region around the burning shell cools, and the hydrogen burning shell cannot reignite.

   \begin{figure}[t]
   \centering
   \vspace{0.5cm}
   \includegraphics[scale=0.55,angle=0]{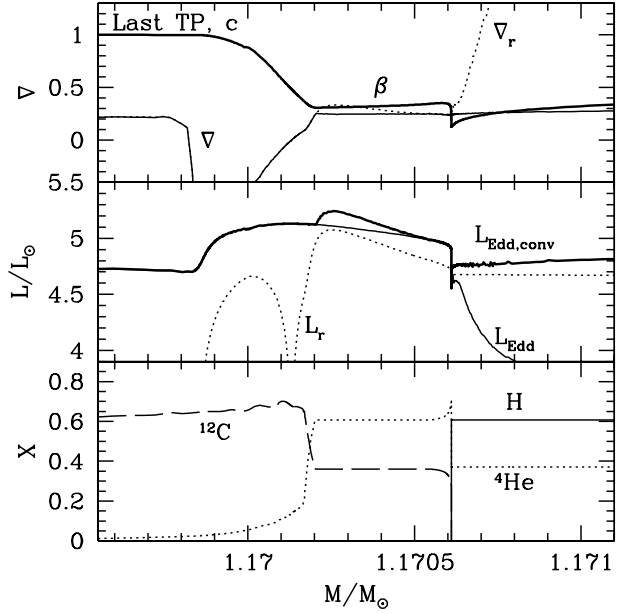}
   \caption{
    Relevant structure and composition parameters for the model labeled
$c$ in Fig\ref{fig:lastlumconv}. Upper panel: actual gradient (thin solid), radiative gradient (dotted line) and $\beta$ (thick solid).
Middle panel: Local luminosity values (dotted line), Eddington luminosity (thin solid) and Eddington luminosity corrected with the 
effect of convection (thick solid line). Lower panel: Hydrogen (solid line), Helium (dotted line) and Carbon (dashed line) composition profiles.}
 \label{fig:lednodiv}
   \end{figure}

\begin{figure}[t]
   \centering
   \vspace{0.5cm}
   \includegraphics[scale=0.55,angle=0]{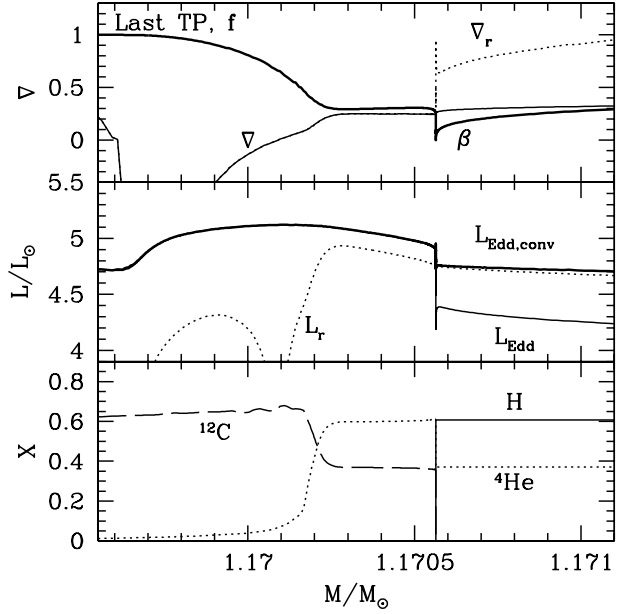}
   \caption{
    Relevant structure and composition parameters for the model labeled
$f$ in Fig.\ref{fig:lastlumconv}. Upper panel: actual gradient (thin solid), radiative gradient (dotted line) and $\beta$ (thick solid).
Middle panel: Local luminosity values (dotted line), Eddington luminosity (thin solid) and Eddington luminosity corrected with the 
effect of convection (thick solid line). Lower panel: Hydrogen (solid line), Helium (dotted line) and Carbon (dashed line) composition profiles.}
\label{fig:leddiv}
   \end{figure}

Fig.~\ref{fig:lednodiv} shows a section of the mass profile of the star near
the BCE during the last converging thermal pulse (at time labelled $\rm{c}$ in Fig.~\ref{fig:lastlumconv}).
Fig.~\ref{fig:leddiv} shows the same as Fig.~\ref{fig:lednodiv} for the time labelled ${\rm f}$, that
is, for our last converged model. 
The upper panel of both figures represent the actual gradient, $\nabla$ (thin solid line),
 and the radiative gradient, $\nabla_{\mathrm {r}}$ (dotted line). Convection dominates the energy transport in the regions where $\nabla_{\mathrm {r}} \gg \nabla$.
The thick solid line represents the value of $\beta=\frac{P_{\mathrm {r}}}{P}$.
As we can see $\beta$ approaches zero at the bottom of the convective envelope.
The middle panels represent the local luminosity, $L_{\mathrm {r}}$ (dashed line), 
the Eddington luminosity computed for radiative regions (thin solid line) and the Eddington luminosity computed for convective regions
(thick solid line), that is, multiplied
by the factor $\frac{\nabla_{\mathrm {r}}}{\nabla}$, $L_{\rm Edd,conv}$ --in red. In the convective region,
the local luminosity could be higher than the  Eddington luminosity, but lower than  $L_{\rm Edd,conv}$, as we have justified above.
However, in the last converged models (see Fig.~\ref{fig:leddiv}), $L_{\mathrm {r}} \sim L_{\rm Edd,conv}$ and $\beta \sim 0$ in the convective zones and we encounter the convergence problem. 
Note that the instability is actually occuring in a very narrow region of about
$1.5 \times 10^{-3} M_{\odot}$ near the base of the convective envelope.
%The cyan dotted line represents the energy release {\bf from} H--burning, and the dotted
%magenta line represents the energy {\bf from} He--burning. 

Note that both the hydrogen and the helium burning shells are
active at the time of the evolution labelled c. But in the very last converged model, labelled f, the hydrogen burning shell is completely switched off.
Finally, for a reference, the lower panels represent the abundance profile of H, $^4$He and $^{12}$C.

%%%%%%%%%%%%%%%%%%%%%%%%%%%%%%%%%%%%%%%%%%%%%%%%%
In a summary, our analysis of the convergence limits in our
1D-stellar evolutionary code allows us to explain a number of consequences
that we actually encounter when we perform full evolutionary computations
of super-AGB stars. The code fails to converge when the envelope masses are still
relatively large (even above $2 M_{\odot}$). The divergence, associated
with the increase of luminosity in the intershell region above $L_{\rm Edd}$,
occurs in narrow region near the base of the convective envelope. Failure of convergence is temporarily solved by increasing the
mixing length parameter through an increase in $\alpha$.
This increase ultimately allows an increase
in the actual Eddington luminosity --conveniently modified by a factor
$\frac{\nabla_{\rm rad}}{\nabla}$ and, therefore, convergence for higher
values of luminosity in the intershell region.

\section{Test of the Fe-peak opacity hypothesis}

%%%%%%%%%%%%%%%%%%%%%%%%%%%%%%%%%%%%%%%%%%%%%%%%%%%%%%%%%%%%%%%%%%%%%%%%%

   \begin{figure}
   \centering
%   \vspace{2.0cm}
   \includegraphics[scale=0.50,angle=0]{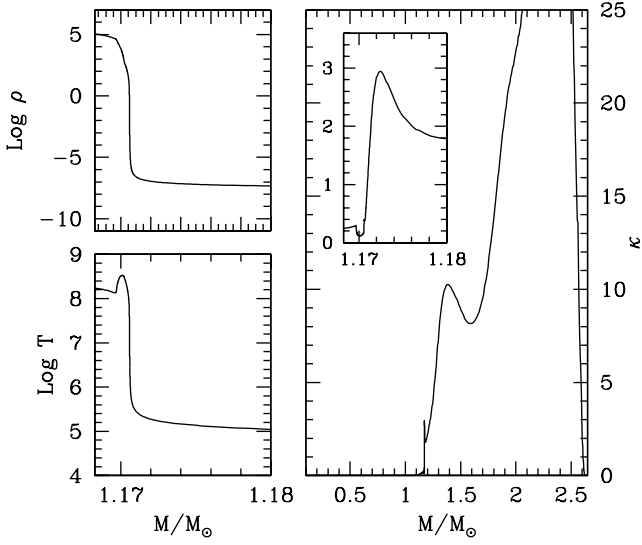}
   \caption{Left upper panel: mass coordinate versus density near the
    $\kappa_{\mathrm{Fe}}$ peak. Left lower panel: mass coordinate versus temperature
    near the $\kappa_{\mathrm{Fe}}$ peak. Right panel. Opacity profile versus mass.
    The insert corresponds to the $\kappa_{\mathrm{Fe}}$ peak.}
   \label{fig:opac1}
   \end{figure}

One possible cause for the instability that
stops the evolution of our model massive AGB and super-AGB stars is
the effect
of the iron opacity peak ($\kappa_{\mathrm{Fe}}$ peak). The $\kappa_{\mathrm{Fe}}$ peak is not a unique feature
in  OPAL tables \citep{igl96}. This feature can also be found in opacities from the Opacity Project (OP). \citet{Seaton04} has a detailed comparison between OPAL and OP opacities.
Typically, the Fe-bump occurs at a  temperature around $2.0\times 10^5$ K, this feature being more pronounced as the metal content increases. The Fe-bump occurs at a higher temperature in OP than in OPAL \citep{Jeffery06}.

\begin{figure}
   \centering
%   \vspace{2.0cm}
   \includegraphics[scale=0.50,angle=0]{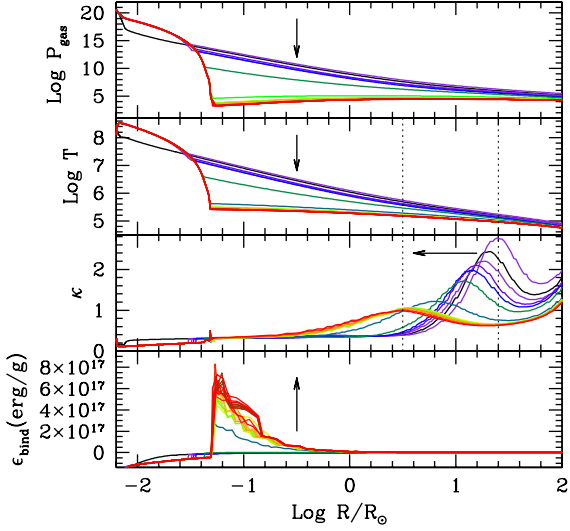}
   \caption{First panel: $Log P_{\rm gas}$; second panel: $Log T$;
    third panel: $\kappa$; fourth panel: $\epsilon_{\mathrm{bind}}$ versus
    the logarithm of radius, $Log R$ near the zone of the instability
    for some selected models prior to the divergence. The arrows
    indicate the direction in which time increases.}
   \label{fig:opac2}
   \end{figure}

The effect of this $\kappa_{\mathrm{Fe}}$ peak in the WR
stars, studied by
 \cite{pet06}, was to decrease the efficiency of energy transport
near the local opacity maximum. For stars more massive than about $15 M_{\odot}$
 and close to the Eddington limit such a decrease in the efficiency of
energy transport would lead to local values of the luminosity above $L_{\mathrm{Edd}}$ 
in the zones just below the opacity peak. 
These authors consider that such situation would lead to a significant
outwards acceleration and the inflation of the stellar envelope.
Even though the stars we are considering here are in the intermediate-mass
range and are in a different evolutionary stage, we
also find the $\kappa_{\mathrm{Fe}}$ peak (see fig~\ref{fig:opac1} and \ref{fig:opac2}) at a
temperature of $t_{\mathrm{peak}}$=$1.6\times 10^5K$, so also close to the value found by \cite{pet06}.
Furthermore our last converged models also show a fast increase in the 
surface radius from $1000 R_{\odot}$ to about $1700 R_{\odot}$ in a 
time interval of 10 years approximately. Therefore, our model star also 
experiences inflation.  \cite{gra11} have recently studied inflation in Wolf-Rayet 
stars and luminous blue variables and established a relation with the topology 
of the $\kappa_{\mathrm{Fe}}$ peak.

In our case the $\kappa_{\mathrm{Fe}}$ peak develops after the helium flash and advances
inwards, both in mass and radius coordinates, as we can see in panel 3 of Fig.~\ref{fig:opac2}.
In this figure we can see the profiles of temperature, gas pressure, opacity and binding energy  
versus the logarithm of the radius, for some selected models between the last helium flash and the 
time of code divergence.
As we have explained, the opacity peak is associated to a certain temperature $T_{\mathrm{peak}}$. This means that an advance
inwards of the $\kappa_{\mathrm{Fe}}$ peak is directly connected to an advance inwards of $T_{\mathrm{peak}}$  and,
therefore, to the cooling and expansion of the affected zone of the star.
The existence of $\kappa_{\mathrm{Fe}}$ peak reduces drastically the efficiency of
energy transport and thus favours the trapping of the energy generated 
at the helium-burning shell. This implies a local increase of internal energy $u_{\mathrm{int}}$ and, as 
can be seen in panel 4 of Fig.~\ref{fig:opac2} an
increase in binding energy $\epsilon_{\mathrm{bind}}$, where $\epsilon_{\mathrm{bind}}=\epsilon_{\mathrm{grav}}+u_{\mathrm{int}}$, and 
$\epsilon_{\mathrm{grav}}$ is the gravitational energy.
$u_{\mathrm{int}}$ increases
at a zone that is expanding while cooling is caused by recombination, and this 
process is fed-back, that is, more cooling favours more recombination that implies more energy for 
expansion and cooling and so on. 
%%%%%%%%%%%%%%%%%%%%%%%%%%%%%%%%%%%%%%%%%%%%%%%%%%%%%%%%%%%%%%%%%%%%%%%%%%%%%%%%%%%%%%%%%%%%%%%%%%%%%

%%%%%%%%%%%%%%%%%%%%%%%%%%%%%%%%%%%%%%%%%%%%%%%%%%%%%%%%%%%%%%%%%%%%%%%%%%%%%%%%%%%%%%%%%%%%%%%%%%%%%%%
Eventually the densities between the helium-burning shell and the BCE decrease so much and so fast that $P_{\rm gas}$ shows an 
inversion, that is, an increase with increasing radius that we can see in the first panel 
of Fig.~{\ref{fig:opac2}. This can be identified as the start of the instability that finally leads to 
the crash of the code. The fact that the instability is associated to a decrease in the 
efficiency in energy transport is supported by the fact that the first model
presenting $P_{\mathrm{gas}}$ inversion corresponds to the time at which the inner convective
shell disappears.

This explanation is consistent with section 2. 
On one hand, the zone of instability identified as the zone of reduced density and $P_{\mathrm{gas}}$ --and even
$P_{\mathrm{gas}}$ inversion, coincides in mass with the zone of instability where $\beta$ tends to zero. Furthermore,
the peak in binding energy also coincides with the zone where $L_{\mathrm {r}}>L_{\mathrm{Edd,conv}}$.
On the other hand, the instability described in this section can be delayed, not only by artificially removing 
the opacity peak $\kappa_{\mathrm{Fe}}$, but also by increasing the $\alpha$ parameter. Both increasing $\alpha$ or decreasing $\kappa$
would allow increasing the efficiency of energy transport and the effect of blocking of energy would be
milder. 
%%%%%%%%%%%%%%%%%%%%%%%%%%%%%%%%%%%%%%%%%%%%%%%%%%%%%%%%%%%%%%%%%%%%%%%%%%

The local maximum in the opacity $\kappa_{\mathrm{Fe}}$ 
might be responsible for a behaviour similar to the one caused by the
$\kappa$-mechanism in
Cepheids and RR-Lyrae. For these pulsating stars, the hydrogen or
helium ionisation regions are responsible for a local opacity increase
and the loss of efficiency in energy transport. This causes an energy excess
below the opacity maximum that, eventually, can be transformed
into work of expansion in the envelope. Once the envelope expands, density and
 temperature decrease and the degree of ionisation also decreases, the star
contracts again and the process repeats.

%In our case, the $\kappa_{\mathrm{Fe}}$ peak develops at the end of a thermal pulse, when the envelope mass is bout 2$M_{\odot}$. It also reduces drastically the efficiency
%of energy transport and, therefore, causes an increase in the luminosity
%below the opacity peak. Furthermore, as $L_{\mathrm{Edd}}$ is proportional
%to $\kappa^{-1}$, the conditions of $L_{\mathrm {r}}>L_{\mathrm{Edd}}$ are favoured and the
%model stars depart from hydrostatic equilibrium.

 \begin{figure}
   \centering
%   \vspace{2.0cm}
   \includegraphics[scale=0.50]{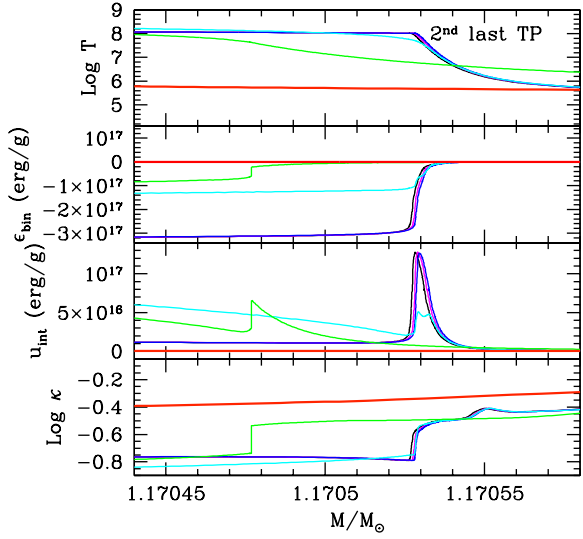}
   \caption{First panel: logarithm of temperature, second panel: specific binding energy
$\epsilon_{\mathrm{bind}}$, third panel: specific internal energy $u_{\mathrm{int}}$ and fourth
panel: logarithm of opacity profiles for a few selected models along the second last thermal pulse,
represented in Fig.~\ref{fig:2lastlumconv}.
   }
\label{fig:binden2}
   \end{figure}

As our code is not hydrodynamic, we cannot follow the evolution further.
Nevertheless it seems reasonable to assume that, as in the massive stars studied by \cite{pet06}
, the energy accumulated below the $\kappa_{\mathrm{Fe}}$ peak
will eventually be released and transformed into work of expansion, so the
(super)-AGB envelopes will also inflate. During this phase of inflation the
mass-loss rates will be correspondingly high and therefore an enhanced superwind is likely
to develop.

\begin{figure}
   \centering
%   \vspace{2.0cm}
   \includegraphics[scale=0.50]{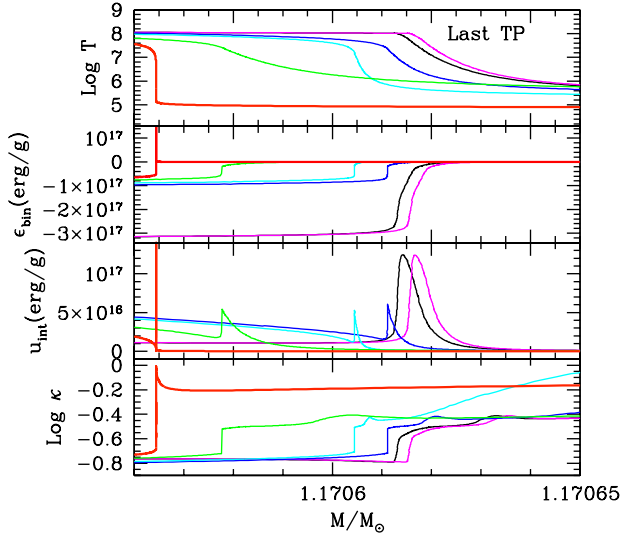}
   \caption{First panel: logarithm of temperature, second panel: specific binding energy
$\epsilon_{\mathrm{bind}}$, third panel: specific internal energy $u_{\mathrm{int}}$ and fourth
panel: logarithm of opacity profiles for a few selected models along the last thermal pulse,
represented in Fig.~\ref{fig:lastlumconv}.
   }
\label{fig:binden}
   \end{figure}
We have tested the hypothesis of the $\kappa_{\mathrm{Fe}}$ peak being
responsible for the divergence of our code by artificially removing this peak.
We have used fig~\ref{fig:opac1} and \ref{fig:opac2} and inspected the data
files of
the structure profiles of the models prior to divergence. Using this information we have imposed
that the opacity for temperatures between $1.5 \times 10^5$ K and $3 \times 10^5$ K and densities
between $10^{-11}$ and $5\times 10^{-7} \mathrm{g/cm^3}$ must have a constant
value equal to the opacity at $3 \times 10^5$ K and density $5\times 10^{-7} \mathrm{g/cm^3}$.
By imposing this constant value, we avoid the instability. This suggests the Fe peak is a major reason for this instability,
but we must take into account that the $\kappa$-peak is real and, unless the energy transport is much more efficient than
what we expect from the mixing length theory, the peak does exist and is likely to lead, as we have seen, to the inflation and,
perhaps, even to the disruption of the star.
%{\bf OK we have two choice, either we just say we avoid the instabiliy. Or we mention it crashed at the end, due to the H \& He ionization zone later. Depends on how much detail we want to give out?}

%______________________________________________________________

\section{What will happen to the star?}

 \cite{woo86} emphasise that the instability that affects TP-AGB 
stars with degenerate cores more massive than $0.89 M_{\odot}$ results from the
disappearance of a hydrostatic solution to the stellar structure
equations. Hence they postulate that the resulting hydrodynamic event
will remove the stellar envelope. 

We have estimated the outward velocity of the envelope by comparing the radii at the same mass mesh points between the last few converging models.
We found that almost throughout the whole envelope, the shell velocities exceed the local escape velocities.
But it is important to note that, because the code crashes during the ejection (as seen from the expansion of the radius of the star), the actual velocity during ejection could be even be higher than the one we are estimating. This problem, in fact, should be 
tackled by using hydrodynamical calculations. 
%{\emph What would you think of adding this? I don't like being so drastic about assuring the enevelope would be ejected.) 

We have also approached the problem of the outcome of the instability by considering the binding energy of the envelope.
Fig.~\ref{fig:binden2} shows the profiles of Log T, specific binding energy $\epsilon_{\mathrm{bind}}=u_{\mathrm{int}}+\epsilon_{\mathrm{grav}}$, specific internal energy $u_{\mathrm{int}}$ and opacity Log$\kappa$ for the models of the second last thermal pulse labelled in Fig.~\ref{fig:2lastlumconv}, and using the same color code. 
Fig.~\ref{fig:binden} shows the same for some selected models of the last thermal pulse, as labelled in Fig.~\ref{fig:lastlumconv}. The specific internal energy $u_{\mathrm{int}}$ has been computed taking into account the terms 
associated to the ideal gas, to radiation, to ionization and dissociation and to electron degeneracy, as in 
\cite{han94}. Wherever
$\epsilon_{\mathrm{bind}}$ has a high negative value the stellar shells are strongly bound and wherever the binding energy approaches zero, the shells are loosely bound. Positive values of the binding energy correspond to the case in which the shells would have enough energy to escape, which can be seen in Fig.~\ref{fig:opac2} and Fig.~\ref{fig:binden}.
The positive values in $\epsilon_{\mathrm{bind}}$ are located between the top of the helium-burning shell ($1.1700 M_{\odot}$ approximately), and with the base of the convective envelope ($1.1706 M_{\odot}$ approximately) in Fig.~\ref{fig:binden}.

The most representative features of the internal energy profiles (third panels of fig~\ref{fig:binden2} and \ref{fig:binden}) are the the local maxima 
that correspond to the zones where energy is accumulated due to the $\kappa_{\mathrm{Fe}}$ peak, as 
we have explained in the former section. As we can see, these peaks practically coincide in mass point with the drop in temperature at the base of the convective envelope and the jump in $\epsilon_{\mathrm{bind}}$.

The second last thermal pulse is characterised by the fact that the mass point 
of the base of the convective envelope and, therefore, also the $u_{\mathrm{int}}$ peaks and $\epsilon_{\mathrm{bind}}$ and $\kappa$ peaks are located practically at the same mass point before the helium flash and until the inner convective shell disappears. Along the last thermal pulse, though, these masses reach values closer to the centre. Finally, when the base of convection reaches the mass point $M=1.17056 M_{\odot}$ and temperatures between 0.1 MK and 1.5 MK, the opacity peak narrows but reaches a diverging value. Due to this  $u_{\mathrm{int}}$ diverges, and so does $\epsilon_{\mathrm{bind}}$.
This is the point at which the it is not possible to proceed further with our calculations.

Another important feature that distinguishes the second last thermal pulse from the last one is 
the variation of the surface radius (see fig~\ref{fig:2lastlumconv} and 
\ref{fig:lastlumconv}). The second last thermal pulse behaves in a standard way. When the helium flash develops the HeBS expand and pushes the HBS that also expands and cools down. This leads to the switching off of the HBS and, without its energy supply gravity dominates and the stellar envelope contracts. The stellar radius only recovers and increases its value when the helium flash is over and the luminosity provided by hydrogen burning is again higher than the luminosity due to helium burning. On the other hand, during the last thermal pulse the
stellar radius remains practically constant. Therefore the energy provided by helium burning is not absorbed in the upper HeBS and intershell region. Instead, it is sufficient to sustain the star against gravity. This situation is stable
as long as the inner convective shell is present. But once this disappears 
while helium burning is still strongly active, a runaway process occurs that leads to the increase of the stellar radius on dynamical time scales and, eventually, to the instability described in this work and to the divergence of the calculations.

From this simple approach, we can derive the same conclusion as \citet{woo86}, that is, the envelope will be ejected shortly after the instability is 
reached. In any case, we would like to point out the limitations of our hydrostatic approach to a problem that would be better treated with 
hydrodynamical codes. Besides, it is also important to realise that the envelope masses we are considering in this work -$\lesssim 2 M_{\odot}$, are 
much higher than the ones considered in \citet{woo86}. Therefore, the amount of work necessary to eject completely the stellar envelopes of massive AGB stars is much higher and, even though our panels showing the binding energy profiles
seem to point to the disruption of the star,
we cannot discard an eventual fallback of the envelopes. 
We have calculated the envelope binding energy for the last computed models as \cite{Meng2008}, and obtained a result of $-3x10^{45}$ erg. This value is negative and, therefore technically the envelope still remains bound. But this value should be considered in perspective. First, the total binding energy of the star is of the order of $10^{49}$. Therefore the envelope mass (about half the mass of the star at this point) has only between $10^{-4}$ and $10^{-3}$ of the total binding energy, so in practice the envelope is very loosely bound to the core.
Second, we can see that the binding energy becomes less negative during the
last computed models. Thus the envelope becomes less bound with
time, until the evolution stops and we cannot proceed further, but this trend
seems to be a good hint of the future behaviour of the star.
A hydrodynamic code would be more suitable to accurately model this ejection process. 

If our estimates showing that most of the envelope (about $\lesssim 2 M_\odot$) will be ejected are correct, the outcomes of our model stars might be relatively massive planetary nebulae,
with some hydrogen-rich matter surrounding a central star of around  $1 M_\odot$. However, it is not clear for how long this post--ejection system could be observed as planetary nebula
because it could move very quickly across the HR-diagram to the white dwarf cooling track \citep{Vas94}. The evolution of a  $1 M_\odot$ post-AGB remnant may be too fast to light a massive PN, and even if all massive AGB stars undergo the ejection mechanism, they are unlikely to be observed as a PN with a massive nebula.
Yet, it is interesting to note that the observational counterpart of such a system could indeed exist.  Planetary nebula N66 contains about $0.6 M_\odot$ of hydrogen-rich matter in the nebula, and the luminosity of the post-AGB star corresponds to a core mass of about $1.2 M_\odot$ \citep{ham03}. 
Indeed, \citet{Vas96} has already suggested that this planetary nebula
could be caused by the radiation ejection mechanism. The very low frequency of N66-like PN with a large nebula could be consistent with the fact that not all AGB stars going through such an ejection mechanism would be observed as a PN

\section{Discussion and conclusions}

In this work we consider the problem of the instability of massive AGB and super-AGB stars and identify the reasons why stellar evolution codes break when envelope masses are still relatively massive --as massive as $2 M_{\odot}$ for the case of super-AGB stars. 
{If this instability is actually encountered by real stars and if it immediately causes their total disruption then it would affect the expected yields of intermediate-mass stars.
Furthermore, the results of our analysis may have an effect on the determination of the mass limit for the formation of electron-capture supernovae, as the presence of the instability we have described might quench the evolution of the star toward core masses above Chandrasekhar mass.
Such an analysis might be particularly important in the case of extremely metal-poor super-AGB
stars, in which computational time is still a problem and whose final fate determination has been often affected by very simple extrapolations.

The problem of the instability of stars at the end of the AGB phase has been considered by \cite{woo86}. 
These authors realised that, near the maximum of a late helium flash during AGB evolution, local values 
of the radiation pressure became so high that they were enough to balance gravity. 
Under these circumstances the gas pressure tended to zero and the local luminosity ($L_{\mathrm {r}}$) became higher 
than the Eddington luminosity ($L_{\mathrm{Edd}}$). Therefore the model stars departed from hydrostatic equilibrium 
and their evolution was quenched. We have computed the evolution of massive AGB stars and realised that the zones at which $L_{\mathrm {r}}>L_{\mathrm{Edd}}$ occurred were at the base of the convective envelope. Therefore we have generalised the results by \cite{woo86} for the case in which the departure of hydrostatic equilibrium developed in the presence of convection and have generalised an expression of $L_{\mathrm{Edd}}$ for convection-dominated environments.

Opacities have frequently been suggested as responsible for the instability. 
For instance, in the classic works by \cite{bak1965} and \cite{bak1965}, it is justified that 
the conditions for the stability of static, radiative layers in gas spheres only depend on the 
local thermodynamical state of these layers --Baker's one-zone model. Thus, if the constitutive 
relations -- equations of state and Rosseland mean opacities -- are specified, the stability conditions can be evaluated without specifying further properties of the layer. In the case of solar-composition
gas spheres the $\kappa$-mechanism can even work in regions where $\mathrm{H}_2$ dissociation and 
H-recombination are occuring, and such regions are much more extended than those of the second ionization
zone that drives Cephe{\"\i}d pulsations. \cite{wag94} also pointed to H-recombination as responsible
for the termination of the AGB phase.
 
We show that, in the case of AGB and super-AGB models at the late stages of their thermally
pulsing phase a similar instability occurs, but it is caused by the Fe-peak in the opacity at the base of 
the convective envelope. This phenomenon was already described by \cite{pet06} for the case of massive stars. 
As in \cite{woo86}, the radiation pressure completely dominates the gas pressure 
and density at the convective envelope converges to zero.
When this instability occurs depends on the treatment of convection. A more efficient scheme of convection or overshooting could delay or even avoid its occurrence. In particular, the value of the mixing length parameter $\alpha$ determines when the instability occurs. This point is very relevant, as it has recently been suggested that $\alpha$ should vary during the evolution of the star --\cite{mea2007}. Moreover, it is possible that mixing length theory cannot adequately describe the convection of the actual star --\cite{stan2011}.  Alternatively, if we could find some
observational constraints on these models, we might be able to use these to
discriminate between different convection theories or different values of
$\alpha$ for this phase of the evolution.

We have seen that for massive AGB stars or super-AGB stars with zero-age main-sequence masses $7-10 M_\odot$ the instabilities 
occur when the envelope mass is $1-2 M_\odot$. The instability could occur for initial masses as low as 
$2.5 M_\odot$ in the solar metallicity case. The stars encountering the instability in this study are all hot bottom burning AGB stars. Therefore, the envelope ejected should be nitrogen rich. The envelope mass when instability occurs decreases with the masses of the AGB. For low mass AGB stars, whether the instabilities occur depend on the choice of the mixing length parameter $\alpha$ and treatment of convection. It is fair to conclude the instability does occur for massive AGB stars and super-AGB stars unless mixing length theory is fundamentally flawed.

Let us speculate briefly on the behaviour of the stellar material in the
region of the instability. Our estimates of the velocity of the convective envelope are greater than the escape velocity. Furthermore, when we consider the binding energy profiles for our last converged models we can see that binding energy reaches positive values.
This suggests that almost all of the envelope could be ejected. In fact the radius of the whole star increases very rapidly.
It is thus possible that the AGB stars will enter post-AGB phase right after the instability. As to the effects of our analysis
on the computed yields of AGB and super-AGB stars, we can expect that the ejection phase, if it occurs, will be very short compared to the whole TP-(super)AGB phase and that,
at this point, the nuclear reactions do not play an important role. The nucleosynthesis of the AGB stars will be truncated due to the ejection and the yield of the s-process isotopes will be lower compared to the alternative assumptions that thermal pulses will keep occuring.

If ejection does occur after the instability, the TP-(super)AGB evolution would be truncated much earlier than in a
model star that underwent 'normal' evolution. This would prevent the occurrence of electron-captured supernovae. Compared to the result of \citet{GilPons07}, \citet{Sie07} and \citet{Poelarends08}, which deduce the possibility of electron-captured supernovae based on the assumption of regular thermal pulses and extrapolation on core growth rate and mass-loss rate, the actual ranges for the occurrence of electron-captured supernovae could be much narrower in terms of both mass and metallicity. For example, our models show that electron-capture supernovae is very unlikely in solar metallicity. On the other hand, if the instability merely leads to a higher mass-loss rate instead of the ejection of the whole envelope, then electron-capture supernovae may still be possible even though the instability is recurrent. The effect of the instability is certainly a factor that cannot be discarded when considering the possibility of electron-captured supernovae. On the other hand,
at lower metallicities, the opacity jump due to the peak would be reduced because of the lower abundance of iron, thus the instability might be reduced and the ejection occurs much later or not occur at all for lower metallicity models. In such a case, if the instability
described by \cite{wag94} cannot operate either, it might possible to
find a critical maximum metallicity for electron-capture supernovae to occur. Evolution of metal-poor super-AGB stars will be modeled in future work.

As a final note, our hydrostatic code could not model through the whole episode of the instability. Although our estimate shows the envelope is likely to be ejected, there are a few processes that cannot be modelled properly, for instance, the situation when the envelope expands and the density converges to zero at the bottom of the convective envelope. As a result, the density of the envelope is inversed from the bottom and induce the Rayleigh-Taylor instability, that could facilitate mixing of material and energy transport. Thus the instability may be overestimated by our code. However, modelling such a process is beyond the capability of the Monash evolutionary code. Modelling on a 3D hydrodynamic code such as Djeuhty \citep{Dearborn2006} may be needed in order to study this instability properly.

\begin{acknowledgements}
This work was supported by the Australian Research Council through grants DP0877317 and DP1095368. HBL would like to thank Norbert Langer for interesting discussion on instability of massive stars. PGP would like to thank the kindness and hospitality of the SINS group at Monash and Jordi Ortiz Domenech for his constant support. We would like to thank the anonymous referee for suggestions improving our manuscript.
\end{acknowledgements}

% for the bibliography, at the end
\bibliographystyle{aa} % style aa.bst
\bibliography{instab.bib} % your references Yourfile.bib

\end{document}